# A combinatorial guide to phase formation and surface passivation of tungsten titanium oxide prepared by thermal oxidation

Sebastian Siol[a,*], Noémie Ott[a], Michael Stiefel[a], Max Döbeli[b], Patrik Schmutz[a], Lars P.H. Jeurgens[a], Claudia Cancellieri[a]

[a]Empa - Swiss Federal Laboratories for Materials Science and Technology, Überlandstrasse 129, 8600 Dübendorf, Switzerland

[b]ETH Zurich, Ion Beam Physics, Otto-Stern-Weg 5, 8093 Zurich, Switzerland

*E-Mail: Sebastian.Siol@empa.ch

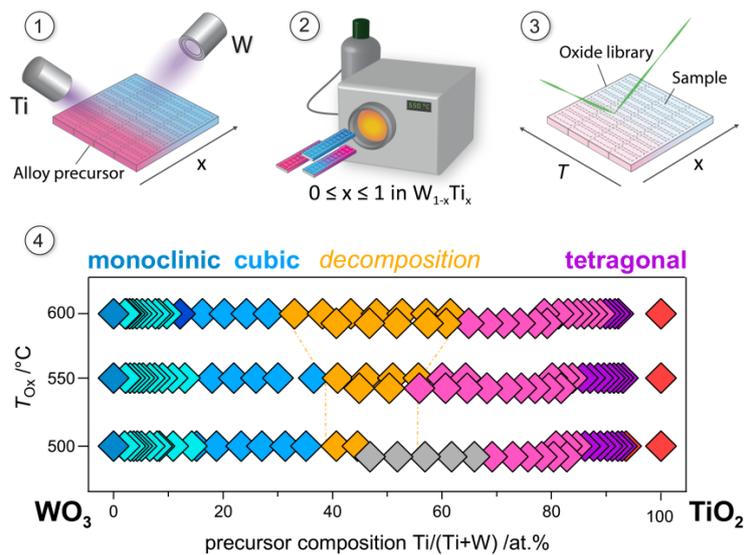

**Graphical abstract:** Using combinatorial thermal oxidation of solid solution $W_{1-x}Ti_x$ precursors combined with bulk and surface analysis mapping we investigate the oxide phase formation and surface passivation of tungsten titanium oxide in the entire compositional range from pure $WO_3$ to $TiO_2$.

**Abstract:**

TiO$_2$ and WO$_3$ are two of the most important earth-abundant electronic materials with applications in countless industries. Recently alloys of WO$_3$ and TiO$_2$ have been investigated leading to improvements of key performance indicators for a variety of applications ranging from photo-electrochemical water splitting to electrochromic smart windows. These positive reports and the complexity of the ternary W-Ti-O phase diagram motivate a comprehensive experimental screening of this phase space. Using combinatorial thermal oxidation of solid solution W$_{1-x}$Ti$_x$ precursors combined with bulk and surface analysis mapping we investigate the oxide phase formation and surface passivation of tungsten titanium oxide in the entire compositional range from pure WO$_3$ to TiO$_2$. The system shows a remarkable structural transition from monoclinic over cubic to tetragonal symmetry with increasing Ti concentration. In addition, a strong Ti surface enrichment is observed for precursor Ti-concentrations in excess of 55 at.%, resulting in the formation of a protective rutile-structured TiO$_2$ surface layer. Despite the structural transitions, the optical properties of the oxide alloys remain largely unaltered demonstrating an independent control of multiple functional properties in W$_{1-x}$Ti$_x$O$_n$. The results from this study provide valuable guidelines for future development of W$_{1-x}$Ti$_x$O$_n$ for electronic and energy applications, but also novel engineering approaches for surface functionalization and additive manufacturing of Ti-based alloys.

## 1. Introduction

TiO$_2$ and WO$_3$ are two of the most important earth-abundant materials for electronic and energy applications.[1,2] WO$_3$ is commonly used in electrochromic devices[2] and gas sensors[2–4], whereas TiO$_2$ is employed across countless industries, including catalysis and energy conversion[1,5–7]. In recent years alloys of WO$_3$ and TiO$_2$ have been investigated[8–10] leading to improvements of key performance indicators for a variety of applications. Alloying WO$_3$ with Ti for instance has been reported to increase both electrochromic contrast and durability in smart windows[11,12], whereas alloying TiO$_2$ with W has been shown to improve catalytic performance[13,14]. Despite these promising results no comprehensive study of the W-Ti-O material system has been performed to date. This is likely rooted in the complexity of its phase diagram. The TiO$_2$-WO$_3$ alloy system is both heterostructural and heterovalent, i.e. the endmember materials exhibit different ground state structures as well as cation-coordination. At normal conditions WO$_3$ crystallizes in the monoclinic structure although other polymorphs, including two higher symmetry tetragonal and cubic phases, have been reported[15,16]. The three most common (out of seven) TiO$_2$ polymorphs are anatase, rutile, and brookite with rutile being the experimentally observed ground state[1,17]. In addition to this high degree of polymorphism both materials are known to form sub-stoichiometric phases, i.e. TiO$_{2-x}$ and WO$_{3-x}$ further increasing the number of potential phases.

The increasing demand for novel oxides with multi-property functionality however motivates the exploration of such increasingly complex material systems. Recent advances in computational materials science have played an important role in exploring previously uncharted phase space[18]. In an effort to accelerate the discovery and development of advanced materials computational data from first principles has been made available for many materials systems in online repositories, including W-Ti-O[19].

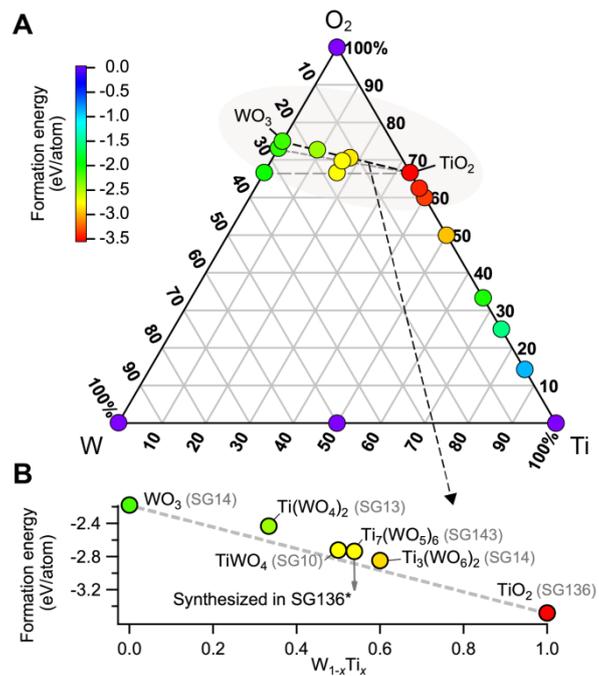

**Figure 1: Calculated formation energies of compounds in the W-Ti-O system. A)** Ternary phase diagram illustrating the formation energies of binary and ternary W-Ti-O compounds. Metastable and unstable compounds are included for the phase space related to this study only. **B)** Convex hull diagram between $WO_3$ and $TiO_2$ as a function of Ti concentration. *Despite being predicted to crystallize in space group (SG) 143 structure $Ti_7(WO_5)_6$ has reportedly been synthesized in SG 136 structure[20]. – data from the "Materials Project" database[19].

**Figure 1** shows enthalpies of formation for the ternary W-Ti-O phase space including a convex hull diagram for the area of interest in this study. A number of metastable compounds are predicted close to the convex hull (Figure 1B). To our knowledge only one of the predicted ternary oxides in this W-Ti-O phase space has been experimentally synthesized[20] which further motivates an experimental screening of this system.

Recently, high-throughput combinatorial experiments have been used for an accelerated screening of synthesis-phase-diagrams in a variety of material systems[21–25]. This type of phase diagram screening is typically performed using non-equilibrium physical vapor deposition (PVD) techniques[25,26], which often results in homogeneous alloy samples with high miscibility[27,28]. While this is beneficial for many applications, the non-equilibrium phase diagrams derived in these studies can be strikingly different from other synthesis techniques[29]; especially for oxidation from metal precursors where the oxide phase formation is often governed by the oxidation kinetics[30,31]. Thermal oxidation is still one of the most relevant industrial processes to produce metal oxides, in particular in sensing and catalysis due to the ability to create complex microstructures[32] with large surface areas, which are hard to realize using PVD approaches. Despite the relevance for industrial applications, few studies combine combinatorial thermal oxidation with state of the art automated characterization[33–35].

In this work we perform a comprehensive screening of the oxide phase formation and passivation of $W_{1-x}Ti_x$ precursors upon thermal oxidation. $W_{1-x}Ti_x$ solid solution precursors are thermally oxidized using a combinatorial synthesis approach combined with automated mapping of bulk and surface properties. In the present study a particular emphasis is put on this combination of bulk and surface characterization. Many combinatorial phase mapping efforts are limited to bulk structural and compositional mapping (e.g. X-ray diffraction (XRD) combined with X-ray fluorescence (XRF)), whereas complete data sets for bulk and surface composition provide valuable information which can inform targeted tailoring of surface properties as a function of alloying concentration and synthesis parameters[35,36]. Using such an approach we demonstrate how oxide phase formation as well as oxide surface passivation change as a function of alloying concentration and processing temperature in tungsten titanium oxide alloys. The system undergoes a structural transition from monoclinic over cubic to tetragonal symmetry with increasing Ti content. In addition, a strong Ti surface enrichment is observed for oxide alloys grown from Ti-rich precursors, leading to the formation of protective $TiO_2$ surface layers on top of homogeneous $W_{1-x}Ti_xO_n$ alloys for Ti alloying concentrations in excess of $x > 0.55$. The results of this study elucidate the phase formation and surface passivation in $W_{1-x}Ti_xO_n$ alloys grown by thermal oxidation, but more importantly demonstrate how combinatorial synthesis coupled with bulk and surface characterization methods can provide insights in the oxidation behavior of previously underexplored material systems.

## 2. Methods
## 2.1 Synthesis

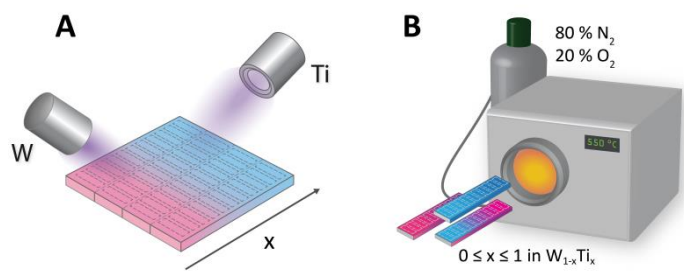

**Figure 2: Combinatorial synthesis of $W_{1-x}Ti_xO_n$ alloys: A)** Metallic precursors are co-sputtered on multiple glass slides at shallow incident angles resulting in sample libraries with compositional gradients **B)** Ensembles of libraries, each covering the entire compositional spectrum, are thermally oxidized subsequently in a tube furnace under constant synthetic air flow (80% $N_2$, 20% $O_2$, ~200 ml/min) at normal pressure

The oxide alloys were produced by thermal oxidation of sputtered thin-film metallic precursors. **Figure 2** illustrates the two main synthesis steps. Metallic thin film precursors were deposited on borosilicate glass slides (12.75 x 51 mm², Eagle XG, Corning) via magnetron sputtering in an AJA 1500-F deposition chamber. The films were sputtered from metallic W and Ti targets (99.95% purity) in pure Ar atmosphere at a process pressure of $5\times10^{-3}$ mbar without intentional heating of the substrate. Compositional gradients were achieved by sputtering off-axis with an incident deposition angle of approximately 30°. No substrate rotation was used during the deposition. In order to avoid silicide formation a 15 nm AlN diffusion barrier was deposited prior to the $W_{1-x}Ti_x$ deposition. Due to the utilization of deposition rate gradients a slight variation in film thickness is inevitable. Precursor film thicknesses were on the order of 120 ± 20 nm. Each deposition run produced 4 sample libraries. Ensembles of alloy libraries covering the entire compositional range were thermally oxidized in a tube furnace under constant synthetic air flow (80% $N_2$, 20% $O_2$, ~200 ml/min) at atmospheric conditions. The temperature was ramped at 10 K/min and held for 15 h at a constant temperature of 500 °C, 550 °C or 600 °C, which ensured complete oxidation of the alloy precursor in the studied compositional and temperature ranges. Real-time synchrotron XRD measurements during oxidation were carried out at the material science beamline (MS-X04SA) at the Swiss Light Source (SLS) in Paul Scherrer Institute (PSI), Villigen, Switzerland[37]. W and Ti films were heated in air while continuously monitoring the oxide peak intensity to determine the temperature threshold for oxide nucleation in either material (see Figure S1). While W already shows nucleation of a monoclinic oxide phase at ~380 °C, Ti requires higher oxidation temperatures above 500°C to form a crystalline oxide. An upper temperature limit of 600 °C was set to minimize stress-induced cracking and delamination of the precursor layers, as well as to suppress the formation of oxygen deficient phases for the W-rich alloys[38]. The annealing time was chosen to be sufficient to fully oxidize Ti-rich precursors. The oxidation rate was found to be higher for W-rich alloys leading to longer the annealing times after complete oxidation. Considering that most synthesis routes are co-determined by thermodynamics and kinetics (e.g. the energy penalties associated with the creation of new interfaces for phase nucleation or decomposition, as well as the atomic mobility in the alloy and oxide systems under study), combinatorial phase screening should be directly performed with the synthesis route of interest to ensure an optimal transferability of the results. We chose thermal oxidation at moderate temperatures over other synthesis methods because of its relevance for industrial processes in the envisioned areas of application for this materials system (e.g. sensing, energy conversion and smart windows).

## 2.2    Characterization

The crystal structure of the precursors as well as oxide films was studied via X-ray diffraction (XRD) using a Bruker D8 diffractometer in Bragg Brentano geometry with Cu Kα radiation and a Ni-filter. X-ray photoelectron spectroscopy (XPS) was conducted using a Physical Electronics (PHI) Quantum 2000 X-ray photoelectron spectrometer featuring monochromatic Al Kα radiation, generated from an electron beam operated at 15 kV and 32.3 W. The energy scale of the instrument was calibrated using Au and Cu reference samples. The analysis was conducted at $1\times10^{-6}$ Pa, with an electron take off angle of 45° and pass energies between 23.50 and 46.95 eV. Surface elemental concentrations were determined in atomic percent using the measured photoelectron peak areas after Shirley background subtraction and the built-in sensitivity factors for calculation. Charge neutralization was performed using a low-energy electron flood gun. The photoemission spectra were aligned using the main component of the C 1s emission at 284.8 eV. X-Ray fluorescence (XRF) measurements were performed in a Bruker M4 Tornado vacuum XRF to determine the bulk composition of the specimen. The combinatorial data sets were processed and analyzed using custom written Igor Pro software routines based on the framework of COMBIgor[39]. The total transmittance (T) was acquired using a UV-Vis-NIR spectrophotometer (Shimadzu UV-3600) equipped with an integrating sphere. The transmittance spectra were fitted using the NanoSCOPS OPTIFIT[40] package (based on the Swanepoel method[41] and the Sellmeier dispersion relation[42]) to estimate the effective refractive index. A Tauc analysis for an indirect transition[43] ($(\alpha h\nu)^{1/2}$) was performed to determine the optical band gap. Rutherford backscattering spectroscopy (RBS) was performed at ETH Zurich on select samples using a 2 MeV He beam. RUMP simulations were used to determine the film composition. The stoichiometry of the films were normalized using W + Ti = 1 for alloyed and pure precursors. The microstructure and chemical composition of selected layers were characterized using conventional and scanning transmission electron microscopy (S/TEM) combined with energy-dispersive X-ray spectroscopy (EDS). The lamellas were prepared using a focused ion beam (FIB) (FEI Helios Nanolab 450s). Imaging was performed using a JEOL 2200FS microscope, operated at 200 kV, fitted with an on-axis bright-field (BF) detector and high-angle annular dark-field (HAADF) detector.

## 3. Results

### 3.1 Structural analysis: X-ray diffraction mapping

Thermal oxidation of the $W_{1-x}Ti_x$ precursor libraries (covering the entire compositional range between pure Ti and pure W) was combined with automated XRD mapping analysis to investigate the oxide phase constitution as function of the alloy composition and the oxidation temperature. **Figure 3** provides an overview of the phase screening process for an oxidation temperature of 600 °C.

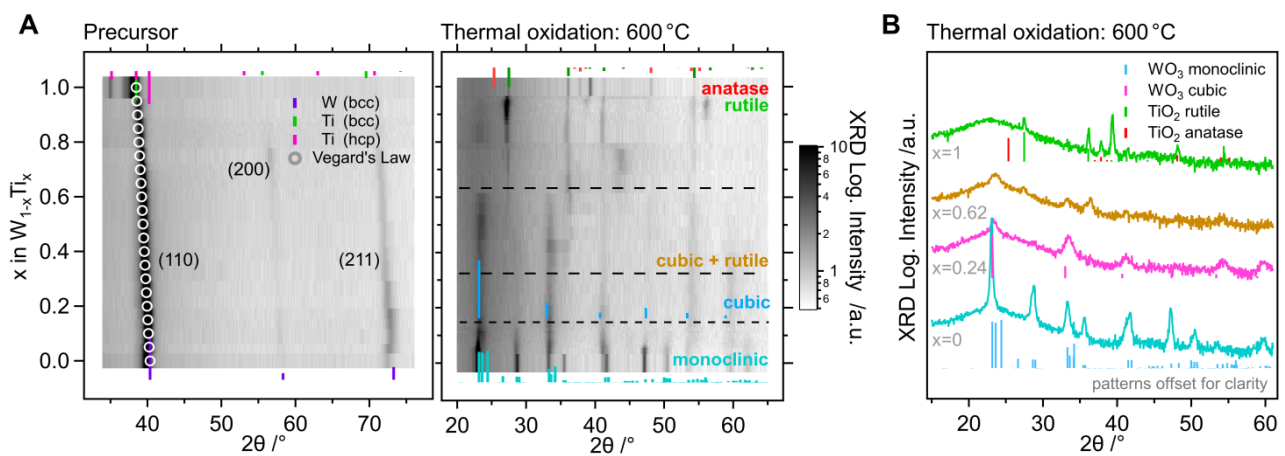

**Figure 3:** Structural characterization of as-deposited $W_{1-x}Ti_x$ precursors and their thermal oxides: A) False color X-ray diffraction maps of the as-deposited $W_{1-x}Ti_x$ precursors and their respective oxide product phases, as obtained after thermal oxidation at 600 °C for 15 h. B) Conventional 2θ - display of few selected oxide diffractograms.

False color plots of the recorded XRD patterns before and after thermal oxidation are plotted as function of the Ti alloying concentration $x$ in **Figure 3A**. With the exception of pure Ti (hexagonal closed pack – HCP) all alloy precursors crystallize in the body centered cubic (BCC) structure. The positions of the (110) and (211) peaks shift as expected according to Vegard's Law with increasing Ti alloying concentration with no visible impurity peaks, indicating the formation of a metastable $W_{1-x}Ti_x$ solid solution for most alloying concentrations $x$.[44] Next the as-deposited alloy precursors were oxidized for 15 hours at a constant temperature in the range of 500 – 600 °C and then again phase screened by XRD. With increasing Ti alloying concentration, the formed oxide undergoes a structural transition from monoclinic to cubic to tetragonal symmetry; a decomposition region, which shows signatures of both cubic and tetragonal phases, occurs at intermediate alloying concentrations. The individual patterns are in good agreement with reference patterns from the literature for monoclinic $WO_3$, cubic $WO_3$, rutile $TiO_2$ and anatase $TiO_2$, respectively. Representative XRD-patterns (in a conventional 2θ-display), showing the principle competing oxide phases formed, are provided in **Figure 3B**. Slight deviations in peak position from the powder diffraction patterns are expected due to the changes in lattice constants upon alloying. The measurement procedure as outlined in Figure 3 was repeated for 3 different thermal oxidation temperatures, i.e. 500 °C, 550 °C and 600 °C. The complete results from this analysis are provided in the supporting information (**Figure S2**). While the same structural transition is observed for all processing temperatures, lower oxidation temperatures lead to overall reduction in crystallinity and a narrower decomposition region with improved solubility for both W- and Ti-rich alloys. In addition to a lower crystallinity observed for lower oxidation temperatures, nucleation of larger grains appears to be suppressed (broader peaks in the XRD pattern) at intermediate alloying concentration due to structural competition. Investigations of the relative position of the main XRD peaks of the oxide product phases as function of the alloying concentration show a discontinuity in the evolution of the $d$-spacing indicating solubility limits of ~10 at.% Ti in monoclinic $(W,Ti)O_{3-y}$ as well as ~10 at.% W in rutile $(Ti,W)O_{2-y}$ (see **Figure S3**).

## 3.2 Microstructure analysis: (S)TEM and SAED

Due to the low crystallinity of many of the grown oxide phases, the assignment of the crystallographic phases from XRD-data alone is approximate and does not adequately capture potential impurity phases, as well as vertical in-homogeneities in the oxide films. To con-firm the results of the XRD screening and reveal possible in-homogeneities, cross-sectional scanning transmission electron microscopy (STEM) analysis was performed on selected oxide samples.

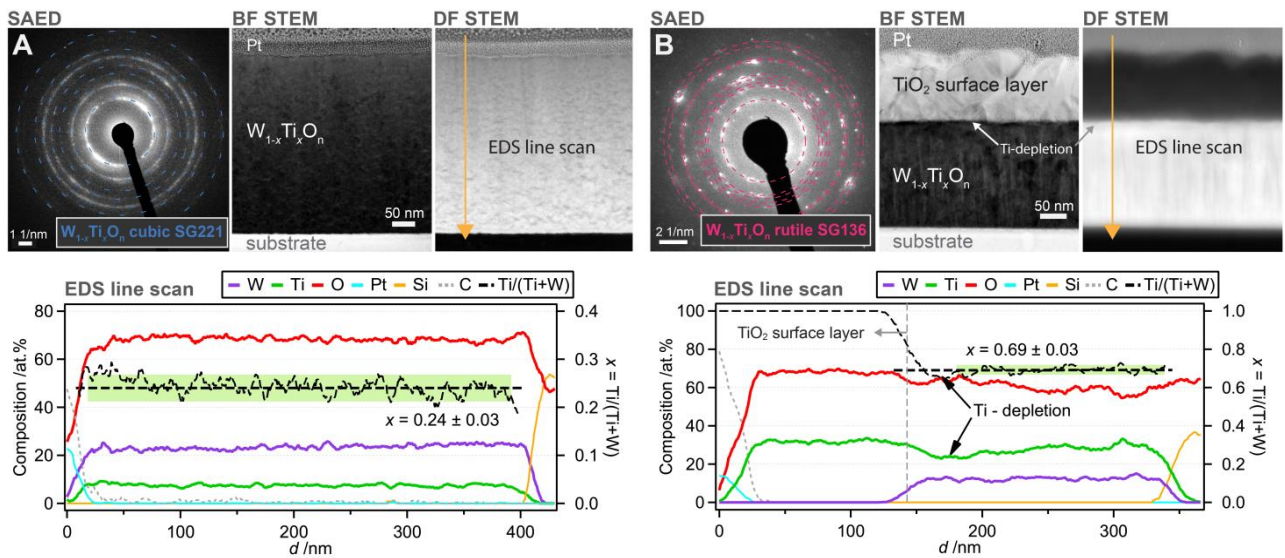

**Figure 4: Microstructure of selected oxide specimens in cross-sectional view:** Shown are selected area electron diffraction (SAED), bright field (BF) and dark (DF) field STEM micrographs as well as electron EDS line scans of oxides with nominal precursor compositions of $W_{0.74}Ti_{0.26}$ (**A**) and $W_{0.28}Ti_{0.72}$ (**B**), oxidized at 550 °C. The oxide grown from $W_{0.74}Ti_{0.26}$ crystallizes in the cubic (SG221) structure. The oxide grown from $W_{0.28}Ti_{0.72}$ consists of a $W_{1-x}Ti_xO_n$ oxide alloy and a $TiO_2$ surface layer both of which crystallize in the tetragonal rutile (SG136) structure. Only minor composition fluctuations are observed in either $W_{1-x}Ti_xO_n$ alloy.

**Figure 4** shows selected area electron diffraction (SAED), bright field (BF) and dark (DF) field STEM micrographs, as well as EDS line scans, of oxides grown at 550 °C with nominal precursor compositions of $W_{0.74}Ti_{0.26}$ (**Figure 4A**) and $W_{0.28}Ti_{0.72}$ (**Figure 4B**). The SAED analysis of the oxidized $W_{0.74}Ti_{0.26}$ alloy indicates that the formed oxide crystallizes in a primitive cubic structure (space group (SG) 223). The dark field (DF) STEM and EDS line scan analyses indicate that the oxide phase formed is structurally and compositionally rather homogenous and possesses an average composition of $W_{0.76}Ti_{0.24}O_{2.2}$. Minor short-range fluctuations of the cation ratio in the measured EDS line scans - despite being within the experimental margin of error - could indicate an initial spinodal decomposition of the formed oxide phase at the nanometer scale. The oxide formed from the oxidation of the Ti-rich $W_{0.28}Ti_{0.72}$ precursor shows a strikingly different phase formation. The SAED analysis indicates the formation of an oxide phase with a tetragonal rutile structure (SG 136). However, the respective diffraction peaks show a slight asymmetry indicating the presence of two types of crystallites with different lattice parameters. This observation is confirmed by the BF and DF STEM micrographs, which clearly show the presence of two distinct layers.

EDS line scans across the bi-layer structure demonstrate the formation of a $TiO_2$ surface layer of 110 nm thickness on top of a homogeneous (W,Ti)-oxide phase with an average composition of ~$W_{0.31}Ti_{0.69}O_{1.8}$ and an average layer thickness of 170 nm. Notably, this $W_{0.31}Ti_{0.69}O_{1.8}$ layer exhibits a Ti-depleted region at its interface with the $TiO_2$ surface layer (see EDS line scan and DF STEM imaging).

## 3.3 Surface analysis: Photoelectron spectroscopy mapping

XPS mapping before and after oxidation was conducted to elucidate the chemical constitutions at the alloy and oxide surfaces (*note:* the average XPS information depth is < 10 nm). The measurements were performed over the entire alloying composition range for oxidation temperatures of 500 °C, 550 °C and 600 °C.

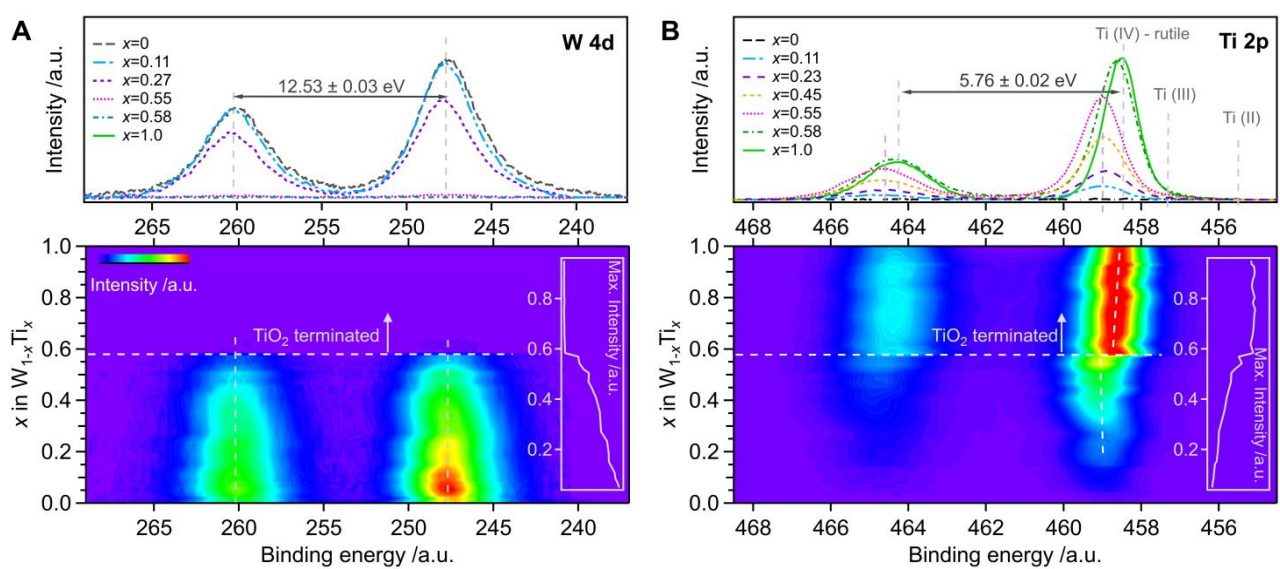

**Figure 5: Photoemission spectroscopy mapping of a $W_{1-x}Ti_x$ precursors thermally oxidized at 500 °C.** Shown are false color XPS maps as function of binding energy and alloying concentration *x*, as well as detailed spectra for selected alloying concentrations. The insets show the maximum intensity over all binding energies at a given *x*. **A)** W 4d core level emission: No peak shift or change in chemical state is observed upon alloying. The signal is completely attenuated for alloying concentrations of x>0.55 **B)** Ti 2p core level emission: Detailed spectra reveal predominantly 4-valent Ti on the oxidized surface. For alloying concentrations of x>0.55 the oxides are $TiO_2$ terminated. The dashed lines represent literature values from the NIST database[45].

**Figure 5** shows the W 4d and Ti 2p core level spectra of oxides produced at 500 °C, highlighting the evolution of the oxide surface chemistry as a function of alloying composition. The W 4d core level is shown, rather than the more common W 4f, since it does not over-lap with the Ti 3p photoemission line. For W, no significant changes are observed as a function of alloying concentration. Both peak-shape and doublet splitting remain largely unaltered in comparison to pure $WO_{3-x}$, indicating the presence of W(VI) throughout. A more detailed analysis of the W 4f emission allows for a fitting of the W(V) vs. W(VI) components (see **Figure S4**). W(V) is often observed in defective $WO_{3-x}$ and directly correlated with the presence of oxygen vacancies[46]. Despite the convolution of the W 4f spectrum with the Ti 3p signal and the limitations of the procedure, a general trend across all oxidation temperatures can be observed, indicating an increasing concentration of 5-valent W with increasing Ti content. The total concentration of W(V) however stays well below 5% for all analyzed samples. Minor fluctuations in the W 4d core level binding energy are present but can be explained by inaccuracies introduced during the C 1s binding energy correction. The W 4d intensity decreases continuously with increasing Ti-concentration, until at a compositional threshold of 55 at.% Ti the detected W signal disappears abruptly. The Ti 2p signal on the other hand (**Figure 5B**) shows a strong increase in intensity at the same compositional threshold. The Ti 2p peak position and doublet splitting for Ti-rich oxides are in good agreement with literature values for rutile $TiO_2$[45]. At lower alloying concentrations, the Ti 2p signal shifts only slightly in binding energy. The fact that the peak shape, as well as the doublet splitting, remain unaltered indicates the presence of 4-valent Ti in the $W_{1-x}Ti_xO_n$ alloy accompanied by a slight change in doping. A screening of the overall oxygen concentration is in good agreement with the W and Ti chemical state analysis, respectively. The approximate surface oxygen to metal ratio for all investigated oxides alongside compositional data from RBS is shown in **Figure S5**.

## 3.4 Optical spectroscopy

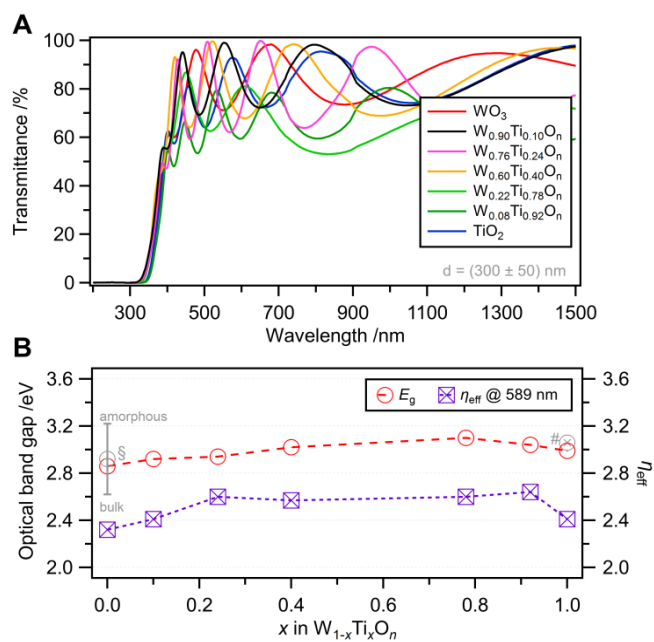

**Figure 6: Optical properties of $W_{1-x}Ti_xO_n$ thermally oxidized at 600 °C: A)** Transmission spectra for various $W_{1-x}Ti_xO_n$ alloys oxidized at 600 °C. **B)** Optical band gaps and effective refraction index $\eta_{eff}$ as function of the alloying concentration. Literature values are provided for reference: §[2], #[47].

$W_{1-x}Ti_xO_n$ alloys are particularly interesting for sustainability and energy conversion technologies. In both of these applications the optical properties are of critical importance. The optical band gap of multiple oxidized alloy precursors was investigated using UV-Vis transmission spectroscopy (**Figure 6**). All samples showed high transparency in the visible range of the spectrum. **Figure 6A** shows transmission spectra of different samples oxidized at 600 °C. Surprisingly, the formation of the bi-layer oxide structure does not strongly affect the optical transmission. By fitting the transmission spectrum after correcting for the substrate absorption the optical band gaps as well as the effective refractive index could be determined (**Figure 6B**). The optical band gap remains largely unaltered with a slight bowing to higher energies for alloying of Ti in $WO_{3-x}$ as well as W in $TiO_{2-x}$. While this change can be due to a change in band structure, it can also be attributed to a general decrease in crystallinity and the resulting smaller average grain sizes for intermediate alloying concentrations, as the band gap of $WO_3$ has been demonstrated to strongly depend on its grain size[2].

## 4. Discussion

### 4.1 Phase evolution and surface passivation of $W_{1-x}Ti_xO_n$ alloys

The combination of XRF-, XRD-bulk measurements as well as XPS-mapping data provides a comprehensive image of the oxide phase formation and preferential oxidation processes in $W_{1-x}Ti_xO_n$ alloys prepared by thermal oxidation (**Figure 7**).

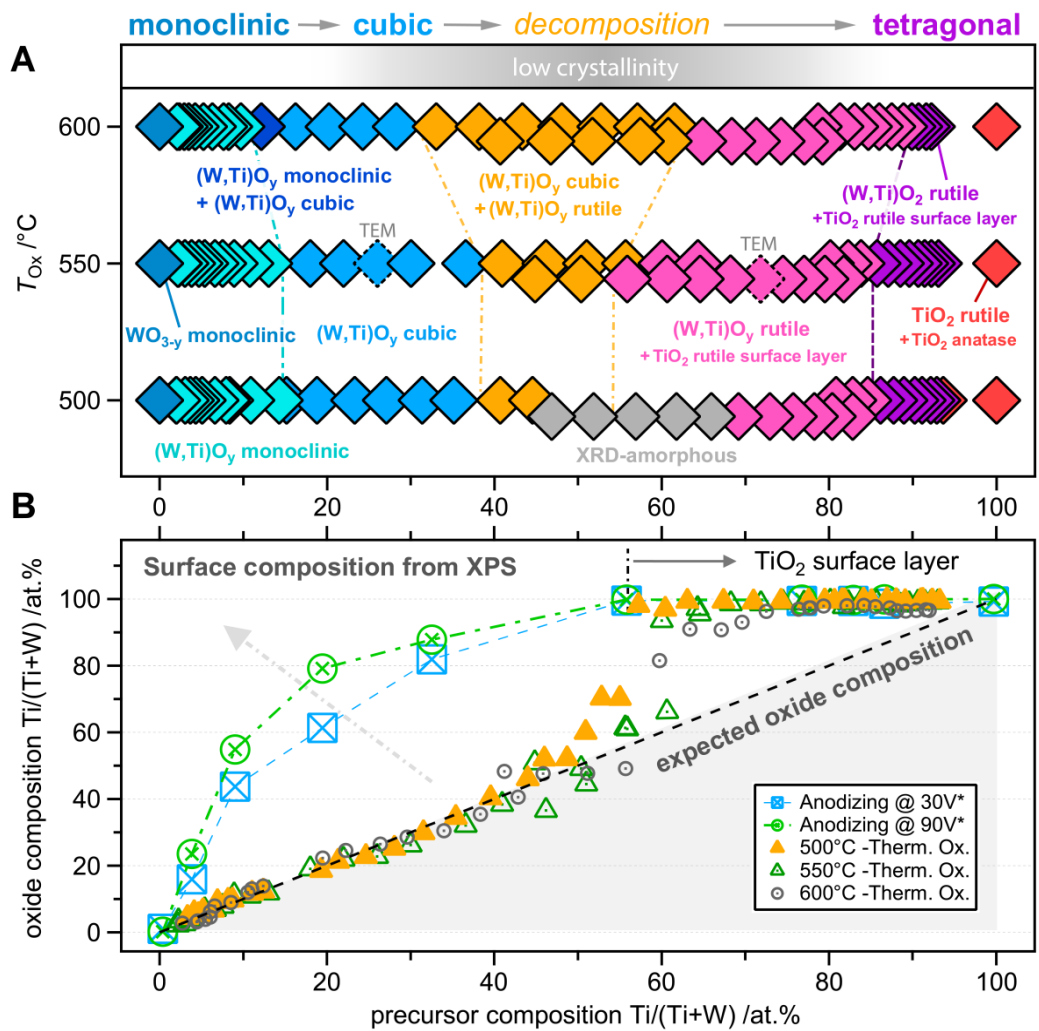

**Figure 7: Structure and surface composition of $W_{1-x}Ti_xO_n$ thermally oxidized at different temperatures. A)** The combinatorial phase map shows the predominant phases from X-ray diffraction measurements. A structural evolution from monoclinic to cubic to tetragonal symmetry is observed with increasing alloying concentration $x$. For intermediate $x$ a decomposition region is observed which becomes more pronounced towards higher oxidation temperatures. **B)** Surface cation enrichment as function of synthesis route and alloying concentration. If the surface composition equals the bulk composition, the data falls on the dotted diagonal with a slope of 1. For alloying concentrations of $x>0.55$ the films are mostly $TiO_2$ terminated. Lower oxidation temperatures provide better passivation. *Ti-enrichment upon barrier anodizing from our previous work is displayed for comparison[48].

**Figure 7A** shows the bulk oxide phase constitution as a function of the alloy precursor composition and the oxidation temperature. For low Ti concentrations monoclinic $WO_{3-x}$ forms. With increasing Ti concentration this phase transforms into a cubic $W_{1-x}Ti_xO_n$ alloy (BCC) of higher symmetry. This is accompanied by a general decrease in crystallinity, which is in good agreement with previous reports in literature[11]. The structure of $WO_3$ can be best described as a network of corner-sharing $WO_6$ octahedra with a defect $ReO_3$ structure[2]. The symmetry of the cubic $ReO_3$ structure is reduced by two types of distortions: a displacement of W from the center of the octahedra and tilting of the $WO_3$ octahedra relative to one another[49]. Multiple polymorphs of lower symmetry have been reported including monoclinic, hexagonal and tetragonal phases. While the monoclinic phase is generally regarded the thermodynamic ground state for $WO_3$, an ordering of the structure can be observed upon annealing at high temperature.[2] But an increase in symmetry can also be achieved by introducing static disorder to the system, either through introduction of oxygen vacancies or doping on the cation site. Depero et al.[49] explained this increase in symmetry from monoclinic to cubic by a connection of cornersharing octahedra via edgesharing. Doping the material with 4-valent Ti leads to a reduction in O-content, and can facilitate this mechanism. In addition to the cubic[11] phase both orthorombic[50] and tetragonal[51] $WO_3$ have been stabilized by Ti-doping.

While the grain size is generally reduced for increased Ti alloying concentrations no crystalline impurity phase is detected by XRD up to a remarkably high Ti alloying concentration of about 35 at.%. A detailed (S)TEM analysis confirms the formation of a homogeneous $W_{1-x}Ti_xO_n$ phase with a cubic structure at 24 at.% Ti (the measured sample is indicated in Figure 7A). Due to a convolution of the specimen's oxygen content with the glass substrate, a quantification of the bulk oxygen content with mapping techniques is challenging. However, the composition from EDS clearly indicates the formation of an oxygen deficient defect phase, which is also confirmed by RBS analysis on samples of similar composition (see **Figure S5**). These results are in very good agreement with the surface characterization from XPS (**Figure 7B**). Up to a Ti alloying content of 40 at.%, the W/Ti atomic ratio at the oxide surface mirrors the bulk composition of the alloy precursor supporting the formation of a homogenous oxide alloy in this composition range.

The oxide phase formation for the Ti-rich side of the phase diagram looks strikingly different. As expected, oxidation of the pure Ti precursors results in the formation of $TiO_2$ with a tetragonal rutile structure. Depending on the oxidation temperature, varying amounts of anatase traces can co-exist in the oxide; higher temperatures lead to a lower anatase phase content in the oxide. This is in good agreement with previous reports in literature. Interestingly, alloying $TiO_2$ with W (as little as 5 at.% is sufficient) stabilizes the rutile structure, even at oxidation temperatures as low as 500 °C. A similar effect has been observed by Aryanpour et al., who found that W-doping enables the formation of $W_{1-x}Ti_xO_2$ in rutile structure, while at the same time increasing the conductivity compared to pristine $TiO_2$[52]. This is particularly interesting for applications in energy conversion and catalysis where the $W_{1-x}Ti_xO_2$ could act as a catalyst support material.[52]

For Ti-concentrations in the range from 40 at.% Ti to approximately 55 at.% Ti, both rutile and cubic signatures is present in the diffractograms indicating a decomposition of the oxide alloys. Starting with this compositional threshold for structural decomposition, a Ti-enrichment on the surface is observed (see Figure 7B). The Ti-enrichment gets more pronounced with higher Ti concentrations and leads to a continuous $TiO_2$ surface layer above a critical Ti alloying content of about 55 at.%. (S)TEM analyses as well as RBS analysis of oxidized Ti-rich alloy precursors with a nominal composition of approximately $W_{0.28}Ti_{0.72}$ confirm the formation of a continuous $TiO_2$ surface layer on top of a homogeneous mixed (Ti,W)-oxide phase. In addition, SAED analysis shows that both oxide layers crystallize in the rutile structure. The surface $TiO_2$ layer increases in thickness with increasing oxidation temperature as evidenced by RBS analysis (**Figure S5**). To elucidate the emergence of this bi-layer structure, high temperature in-situ XRD measurements on Ti-rich alloy precursors where conducted in air (see Figure S6). After an initial peak shift of the metal peak due to O-dissolution, a mixed (W,Ti)-oxide phase with a rutile structure is formed. This rutile phase stays stable up to 800 °C after which decomposition into the endmembers rutile $TiO_{2-x}$ and monoclinic $WO_{3-x}$ occurs. From the experimental findings presented above, a hypothesis for the underlying thermal oxidation mechanism of $W_{1-x}Ti_x$ precursors can be drawn. The onset of oxidation proceeds by the preferential oxidation of Ti from the $W_{1-x}Ti_x$ alloy. As a result, a surface $TiO_2$ layer can develop for Ti alloying concentrations over 55 at.%. The thermal oxidation of Ti metal and Ti-alloys generally follows an oxygen vacancy diffusion mechanism, which is accompanied by a continuous dissolution of O from the developing oxide into the underlying alloy at the wake of the oxide growth front[53]. The interstitially dissolved O can diffuse and redistribute in the underlying alloy until the solubility product for the formation of a $W_{1-x}Ti_xO_n$ solid solution oxide phase is locally exceeded. In this initial stage of oxide nucleation the W/Ti atomic ratio of the nucleating $W_{1-x}Ti_xO_n$ phase will mirror the composition of the partially Ti-depleted alloy precursor. As such, a double-layered $TiO_2/W_{1-x}Ti_xO_n$ oxide structure can develop for Ti alloying concentrations over 55 at.%. Further growth of the $TiO_2$ overlayer then proceeds by the uptake of Ti from the underlying $W_{1-x}Ti_xO_n$ solid solution phase, resulting in a Ti-depletion zone in $W_{1-x}Ti_xO_n$ layer at the interface with the $TiO_2$ top layer. The TiO2 surface layer thickness increases towards higher oxidation temperatures due to the thermally enhanced atomic mobility of Ti in the developing oxide phases (see Figures 4B and S5). The corresponding oxidation mechanism is outlined in Figure S7. Strikingly, the $W_{1-x}Ti_xO_n$ solid solution phase adopts a rutile structure, which should be highly metastable. This indicates that local formation of $WO_3$ and $TiO_2$ (as preferred by bulk thermodynamics) is kinetically hindered due to the limited atomic mobilities of W, Ti and dissolved O in the alloy (for the studied temperature range up to 600 °C)[54]. This result suggests an oxide phase evolution governed by kinetic limitations following Oswald's rule of stages[55] and once more highlights the importance of conducting synthesis phase screenings specifically for the synthesis route of interest.

**Figure 7B** summarizes the effect of preferential oxidation on the evolving oxide surface composition as a function of alloying concentration and oxidation temperature. We recently reported on the formation of $TiO_2$-passivated $W_{1-x}Ti_xO_n$ oxide phases, prepared via barrier anodizing of alloy precursors[48]. Strikingly, the critical Ti alloying content required for the formation of a continuous and protective $TiO_2$ surface layer is similar for both oxidation processes (i.e. anodizing vs. thermal oxidation). However, for Ti alloying contents below this critical Ti concentration, the barrier anodizing process produces a much more pronounced Ti-surface enrichment in the formed oxide phase (Figure 7B)[48]. During barrier anodizing a very high external electric field is applied to accelerate oxide formation on the $W_{1-x}Ti_x$ alloy, far from (local) thermodynamic equilibrium conditions, which apparently results in a relatively high Ti concentration at the developing oxide surface. On the contrary, oxide formation by thermal oxidation proceeds at a much slower rate by ionic diffusion under influence of the chemical potential gradients across the developing oxide layer, much closer to local thermodynamic equilibrium. As a result, the Ti-enrichment at the developing oxide surface during the thermal oxidation process is much less pronounced. While the oxide constitution and surface passivation strongly depend on the alloy precur-sor composition and the oxidation conditions, the optical properties of the material surprisingly remain largely unaltered (Figure 6), which is rooted in the similar refractive indices and band gaps for both $WO_3$ and $TiO_2$. This result is particularly interesting as it demonstrates the ability to independently control multiple materials properties in tungsten titanium oxide.

**4.2 Summary and conclusions**

The results of this study provide a comprehensive overview of the thermal oxidation of $W_{1-x}Ti_x$ alloys, facilitating future materials development and properties tuning in this phase space.

The system shows a strong preferential oxidation of Ti, which results in the formation of a $TiO_2$ surface layer for precursor compositions in excess of 55 at.% Ti. The formation of this surface layer formation is most pronounced for intermediate alloying concentrations and correlates with the occurrence of a phase decomposition in the (W,Ti)-oxide alloys. By comparing the results with our previous work on anodic oxide formation on $W_{1-x}Ti_x$[48], we demonstrate that the degree of surface passivation in $W_{1-x}Ti_xO_n$ can indeed be tailored by changing the oxidation kinetics, a feature that can be used to increase the chemical stability in reactive environments. The oxide bi-layer formation on Ti-rich alloy precursors on the other hand clearly motivates a combination of bulk and surface analysis mapping for future synthesis phase diagram screenings of novel oxide alloys, in particular for the oxidation of metallic precursors or synthesis routes that include annealing steps in reactive atmospheres. Even though none of the theoretically predicted metastable ternary W-Ti-O phases were synthesized (see Figure 1), a remarkable phase transformation from monoclinic over cubic to tetragonal crystal structures is observed with increasing Ti alloying concentrations. On the W-rich side of the phase diagram, Ti-doping leads to an increase in symmetry and the stabilization of a cubic $W_{1-x}Ti_xO_n$ phase. On the $TiO_2$ side doping with W leads to a stabilization of the rutile phase upon inclusion of $WO_2$. While the system undergoes these structural phase transitions, the optical properties remain largely unaltered, which implies that an independent control of multiple functional properties in $W_{1-x}Ti_xO_n$ is possible.

Overall, the results of this work provide a comprehensive guide to phase formation and passivation in $W_{1-x}Ti_xO_n$. In future work, a more detailed analysis of compositional regions of interest in the phase diagram can enable an optimization of the materials properties specific to a targeted application. While the results of this study provide valuable guidelines for future development of $W_{1-x}Ti_xO_n$ for electronic and energy applications, they can also inform other areas of research: For instance, laser surface structuring or additive manufacturing of Ti-based alloys, for which a thorough knowledge of the surface oxide formation is beneficial.

**Acknowledgements**

The authors would like to thank Mirco Chiodi and Yeliz Unutulmazsoy for helpful discussions and assistance with the synchrotron oxidation experiments. We would like to acknowledge the Nanoscale Materials Science laboratory for access to the coating facilities as well as the MS-X04SA beamline at Swiss Light Source (SLS), PSI Villigen for providing beam time for the real-time oxidation experiments. Funding from COST project IZCNZ0-174856 C16.0075, in the COST Action MP1407 (e-MINDS) is gratefully acknowledged.

*Supplementary Information*

# A combinatorial guide to phase formation and surface passivation of tungsten titanium oxide prepared by thermal oxidation

Sebastian Siol[a,*], Noémie Ott[a], Michael Stiefel[a], Max Döbeli[b], Patrik Schmutz[a], Lars P.H. Jeurgens[a], Claudia Cancellieri[a]

[a]Empa - Swiss Federal Laboratories for Materials Science and Technology, Überlandstrasse 129, 8600 Dübendorf, Switzerland

[b]ETH Zurich, Ion Beam Physics, Otto-Stern-Weg 5, 8093 Zurich, Switzerland

*E-Mail: Sebastian.Siol@empa.ch

# 1. Structural analysis

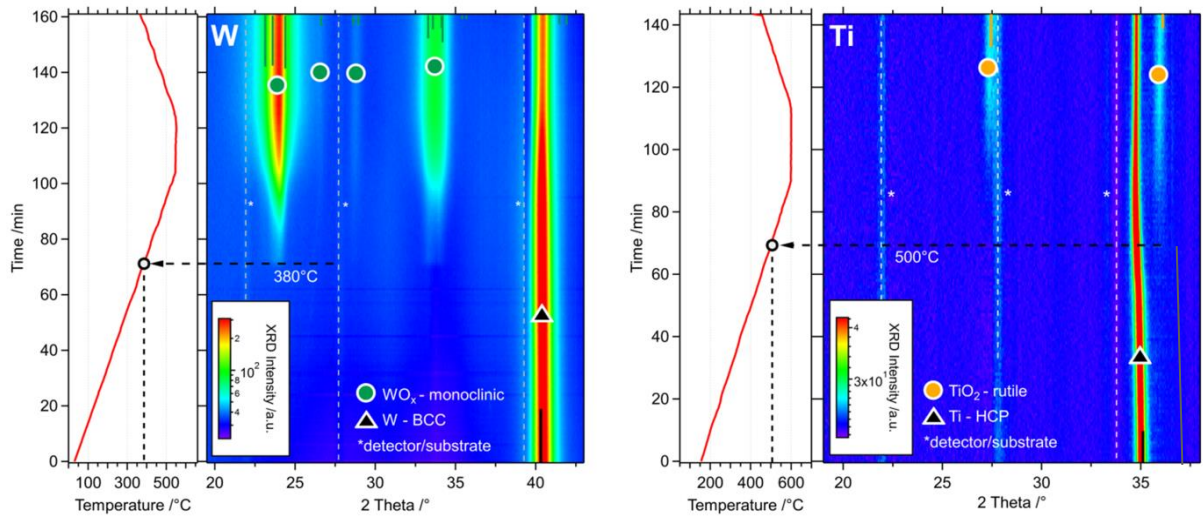

**Figure S1: Real-time synchrotron oxidation experiments on W and Ti:** The graphs show real-time synchrotron XRD-measurements in transmission mode at the PSI SLS beamline. The 2θ-scale was converted to Cu-kα radiation. Thin films of W or Ti were heated in air, while XRD-patterns were measured continuously every minute.

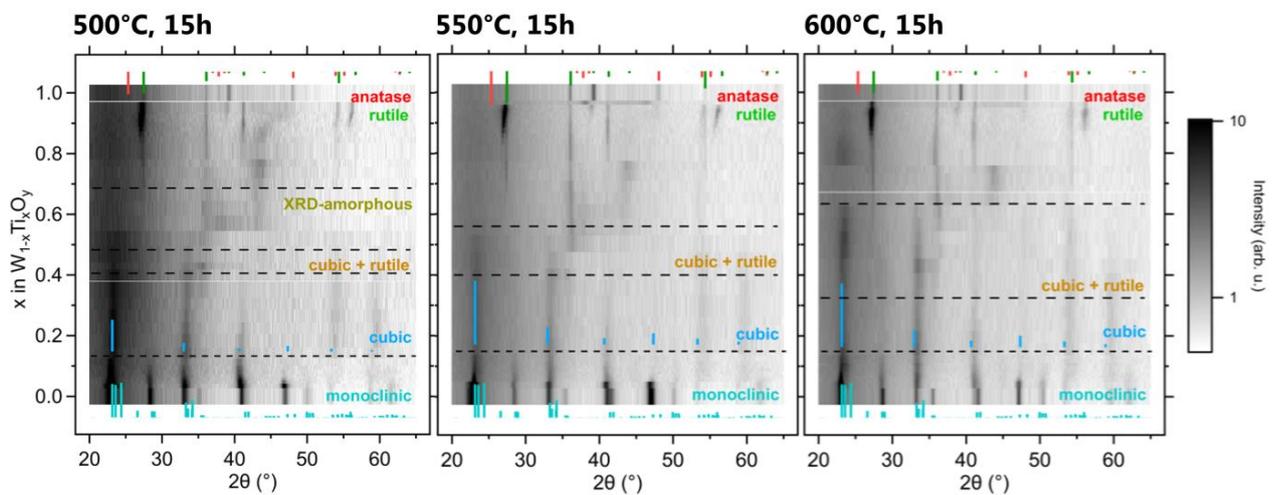

**Figure S2: XRD mapping:** Complete dataset for the structural analysis of the W-Ti precursors after thermal oxidation for 15 h at 500 °C, 550 °C and 600 °C, respectively.

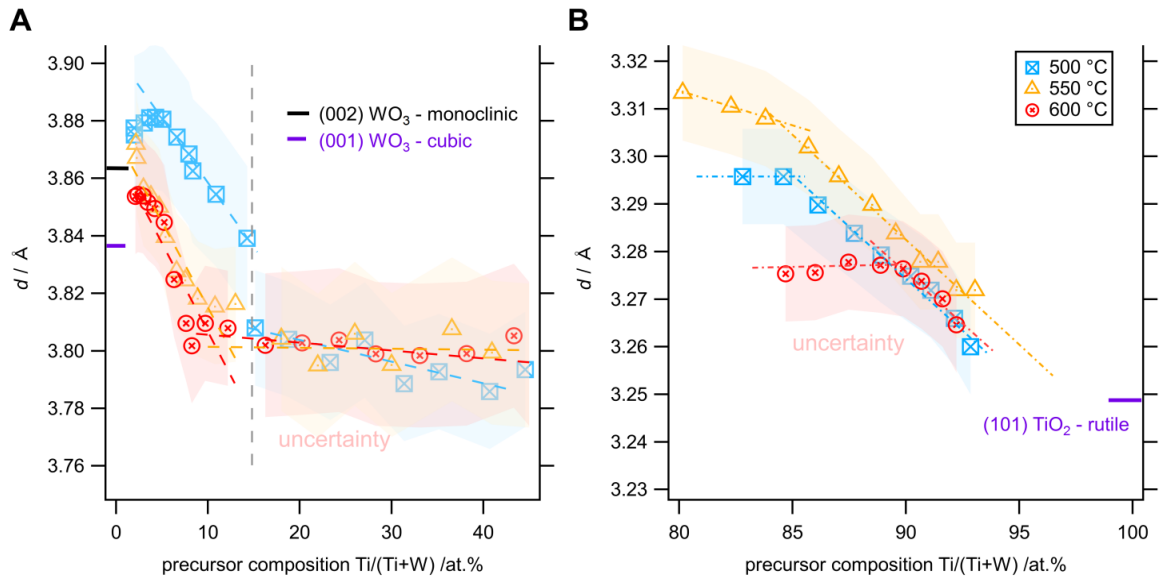

**Figure S3: XRD-peak shift as function of alloying concentration and oxidation temperature:** When plotting the d-spacing corresponding to the main XRD peak discontinuities in the evolution of the d-spacing as a function of alloying concentration can indicate the formation of impurity phases or interstitial defects.

## 2. Surface analysis using XPS

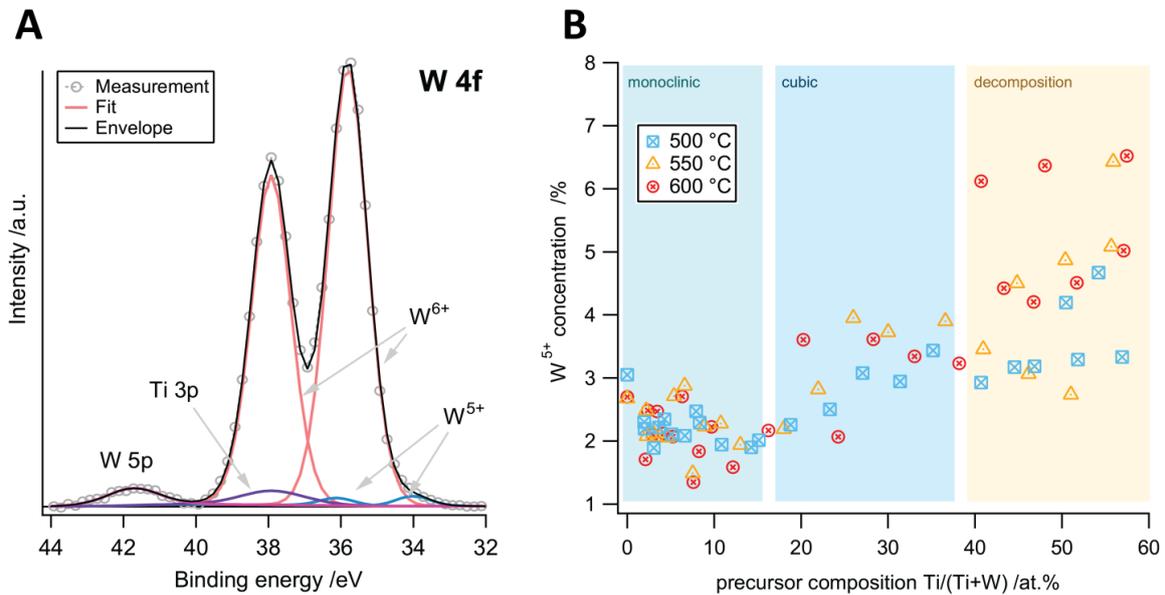

**Figure S4: Chemical state analysis of W:** A) Representative core level emission spectrum of a W-rich oxide alloy. Shown are the W 4f doublets associated with W 6+ and W 5+, respectively. The Ti 3p emission overlaps with the W 4f spectral region. B) $W^{5+}/(W^{5+} + W^{6+})$ ratio as a function of composition and thermal oxidation temperature.

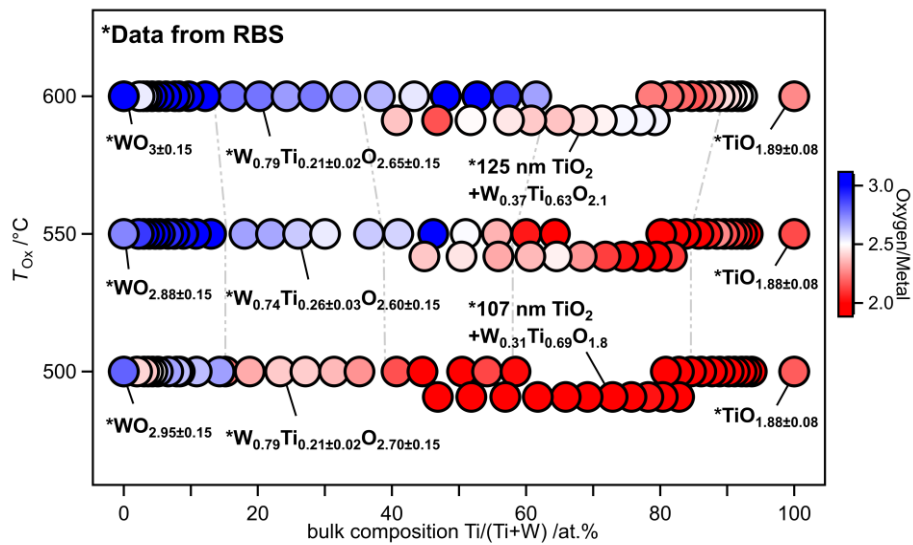

**Figure S5: Oxygen to metal ratio as a function of alloying concentration and oxidation temperature:** The circles represent the surface oxygen to metal ratio measured by XPS. The metal oxide component of the O 1s core level emission was compared to the combined Ti 2p and W 4d emissions weighted by the respective relative sensitivity factors. *Bulk measurements from RBS are shown for comparison.

## 3. Real-time XRD oxidation

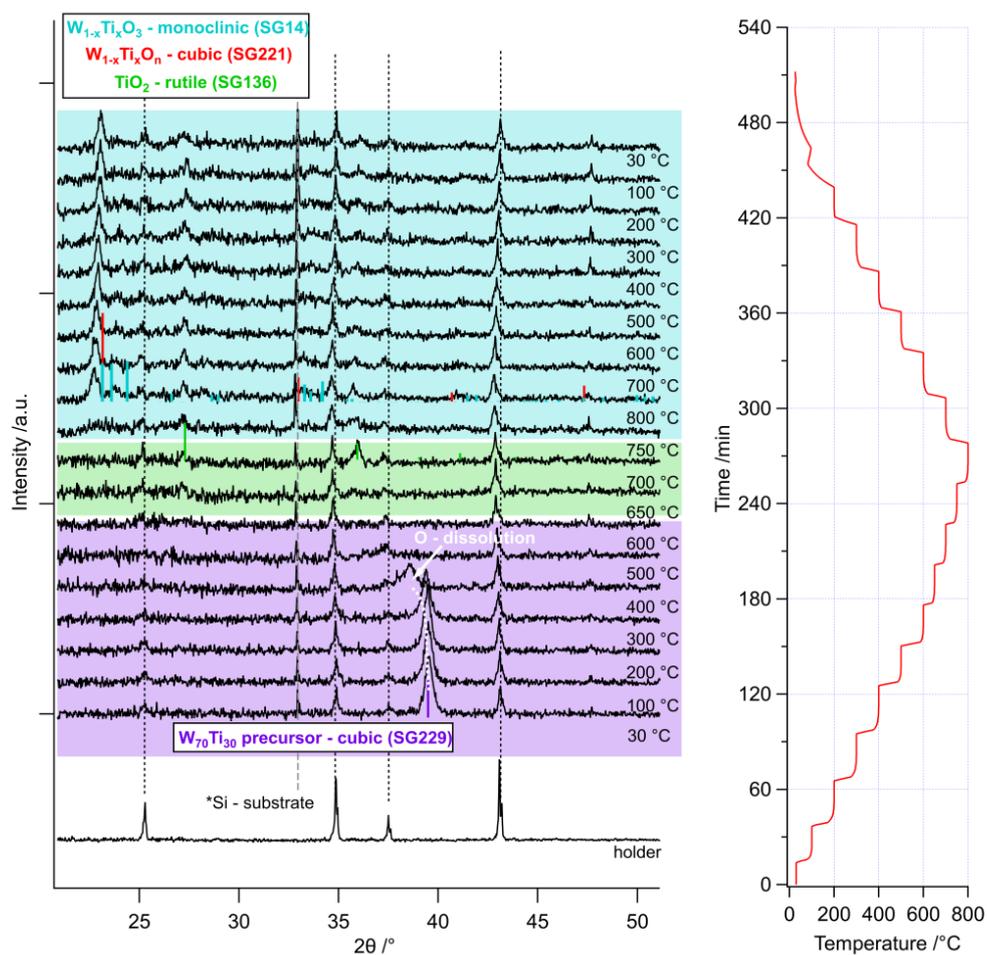

**Figure S6: High temperature in-situ XRD measurements on $W_{30}Ti_{70}$:** A Ti-rich precursor was heated in air while continuously conducting standard XRD measurements. At temperature below 500 °C a shift of the metal precursor peak indicates oxygen dissolution. At higher temperatures of 650 °C a rutile phase is observed which increases in intensity with increasing temperature. At 800 °C the specimen starts to undergo binodal decomposition into a monoclinic tungsten-rich and a rutile Ti-rich phase.

## 4. Thermal oxidation mechanism

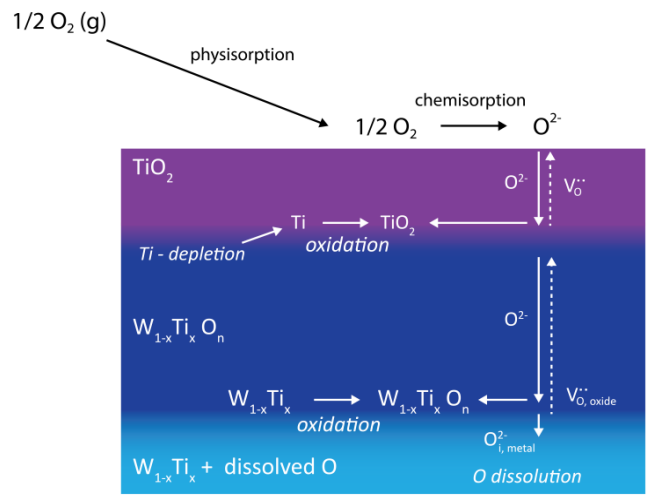

**Figure S7: Schematic illustration of the oxidation process of a Ti-rich $W_{1-x}Ti_x$ alloy:** The oxidation of Ti-rich precursors likely occurs at two reaction fronts. After initial formation of a surface $TiO_2$, O is dissolved in the metal precursor. After reaching a critical O-concentration in the precursor, a mixed oxide is formed. This mixed oxide phase is subsequently depleted of Ti to increase the thickness of the thermodynamically preferred $TiO_2$ surface layer. Multiple separate diffusion processes can control the kinetics of the oxide phase formation. More detailed investigations are necessary to pinpoint the exact oxidation processes resulting in the observed oxide bilayer structure.